\documentclass[aps,prb,preprint,superscriptaddress,nofootinbib,showkeys]{revtex4-2}

\usepackage[T1]{fontenc}
\usepackage[utf8]{inputenc}
\usepackage{amsmath,amssymb}
\usepackage{graphicx}
\usepackage{bm}
\usepackage{comment}
\usepackage{hyperref}

\begin{document}

\title{Anisotropic domain wall velocity profiles in the creep regime: the interplay of chiral damping, stiffness and Dzyaloshinskii--Moriya interaction}

\author{A. Di Pietro}
\email{adriano.dipietro@inrim.it}
\affiliation{Istituto Nazionale di Ricerca Metrologica, Torino, Italy}

\author{A. Magni}
\affiliation{Istituto Nazionale di Ricerca Metrologica, Torino, Italy}

\author{S. Pizzini}
\affiliation{Universit{\`e} Grenoble Alpes, CNRS, Institut N{\'e}el, Grenoble, France}

\author{F. D{\"o}rr}
\affiliation{Department of Physics, Freie Universit{\"a}t Berlin, Berlin, Germany}

\author{Y. Shokr}
\affiliation{Nuclear Radiation Detectors Research and Development Center, Bolu Abant Izzet Baysal University, Bolu, T{\"u}rkiye}
\affiliation{Department of Physics, Freie Universit{\"a}t Berlin, Berlin, Germany}

\author{G. Durin}
\affiliation{Istituto Nazionale di Ricerca Metrologica, Torino, Italy}

\author{S. Tacchi}
\affiliation{Istituto Officina dei Materiali del CNR (CNR-IOM), Sede Secondaria di Perugia, c/o Dipartimento di Fisica e Geologia, Universit{\`a} di Perugia, Perugia, Italy}

\author{M. Madami}
\affiliation{Dipartimento di Fisica e Geologia, Universit{\`a} di Perugia, Perugia, Italy}

\author{G. Carlotti}
\affiliation{Dipartimento di Fisica e Geologia, Universit{\`a} di Perugia, Perugia, Italy}

\author{E. Darwin}
\affiliation{School of Physics and Astronomy, University of Leeds, Leeds, United Kingdom}
\affiliation{Empa, D{\"u}bendorf, Switzerland}

\author{A. J. Huxtable}
\affiliation{School of Physics and Astronomy, University of Leeds, Leeds, United Kingdom}

\author{C. H. Marrows}
\affiliation{School of Physics and Astronomy, University of Leeds, Leeds, United Kingdom}

\author{B. J. Hickey}
\affiliation{School of Physics and Astronomy, University of Leeds, Leeds, United Kingdom}

\author{M. Kuepferling}
\affiliation{Istituto Nazionale di Ricerca Metrologica, Torino, Italy}

\date{\today}

\begin{abstract}
The asymmetric expansion of magnetic bubble domains in ultrathin ferromagnets provides a powerful route to probe the interfacial Dzyaloshinskii--Moriya interaction (DMI). While conventional analyses rely on domain wall velocities measured along selected directions as a function of in-plane field, recent approaches have highlighted the additional insight contained in the angular dependence of the velocity, $v(\theta)$. Here, we develop an extended angular creep model that incorporates both the dispersive domain wall stiffness and a chirality-dependent prefactor associated with chiral damping. This generalization captures the full anisotropic dynamics of domain wall motion around a bubble domain. We show that these contributions significantly modify the angular velocity profile and can lead to features not accessible within existing models. Our results establish a more complete framework for interpreting creep-driven domain expansion and provide improved sensitivity for the quantitative extraction of DMI and chiral dynamical effects.
\end{abstract}

\keywords{racetrack memory, Dzyaloshinskii--Moriya interaction, creep regime, domain-wall stiffness, chiral damping}

\maketitle

\section{Introduction}
Magnetic racetrack memories \cite{parkin_memory_2015,Song2017,blasing_magnetic_2020, gu_three-dimensional_2022} rely on the controlled motion of chiral domain walls (DWs) \cite{Thiaville2012} along nanotracks, where the DW velocity, stability, and reproducibility determine device performance \cite{Vedmedenko2020,Kuepferling2020}. In ultrathin ferromagnetic heterostructures with perpendicular magnetic anisotropy (PMA), these properties are strongly governed by the interfacial Dzyaloshinskii-Moriya interaction (iDMI) \cite{DZYALOSHINSKY1958241,MOR-60,fert_role_1980}, which fixes the DW chirality, stabilizes N\'eel walls \cite{robertson_-plane_2020}, and enables efficient spin-orbit-torque-driven motion \cite{ishikuro_dzyaloshinskii-moriya_2019, bhowmik_deterministic_2015}. Reliable DMI quantification is therefore essential for racetrack-memory materials.
A widely used method to determine the DMI is asymmetric bubble expansion under an out-of-plane driving field and an in-plane magnetic field $H_x$ \cite{Je2013,Shahbazi2019}.  The in-plane field either reinforces or compensates the DMI field depending on the DW magnetization orientation, producing a velocity asymmetry that is commonly used to infer the DMI strength. However, in the thermally activated, low velocity \cite{metaxas_creep_2007} creep regime the DW velocity depends exponentially on the elastic energy scale of the wall. The relevant quantity is therefore not only the equilibrium DW energy density \cite{Lemerle1998,chauve_creep_2000}, but the dispersive DW stiffness, which includes the angular curvature of the energy \cite{lau_disentangling_2018,pellegren_dispersive_2017,Hartmann2019}. As a result, stiffness-induced changes in the creep velocity can be mistaken for changes in the DMI field.
Additional asymmetries can arise from chirality-dependent dynamical prefactors, such as chiral damping \cite{jue_chiral_2016,akosa_phenomenology_2016,akosa_decoupling_2024}. This is particularly relevant for device-oriented DMI metrology, where the same material parameters are used to predict DW motion in confined racetrack geometries. Recent angular bubble-expansion measurements, which resolve the full velocity profile $v(\theta)$, provide a route to access the anisotropic DW energy landscape from a single DW expansion image acquisition \cite{pakam_anisotropic_2024} thus reducing the experimental effort significantly. Yet current angular models rely mainly on orientation-dependent wall energy and neglect stiffness and chirality-dependent damping terms \cite{pakam_anisotropic_2024,di_pietro_effect_2025}.
In this work, we extend the angular velocity framework to include both dispersive DW stiffness and chiral damping. By comparing $v(\theta)$ with and without the stiffness contribution, and by varying the chiral damping term, we clarify how elastic and dissipative effects reshape the apparent DMI-induced velocity asymmetry. Our results show that reliable DMI extraction for racetrack-relevant materials requires separating energetic, elastic, and dissipative contributions to DW motion, rather than assigning the measured asymmetry to DMI alone.

\section{Angular creep model}

\begin{figure}
    \centering
    \includegraphics[clip, trim= 0cm 1cm 0cm 2cm,width=\linewidth]{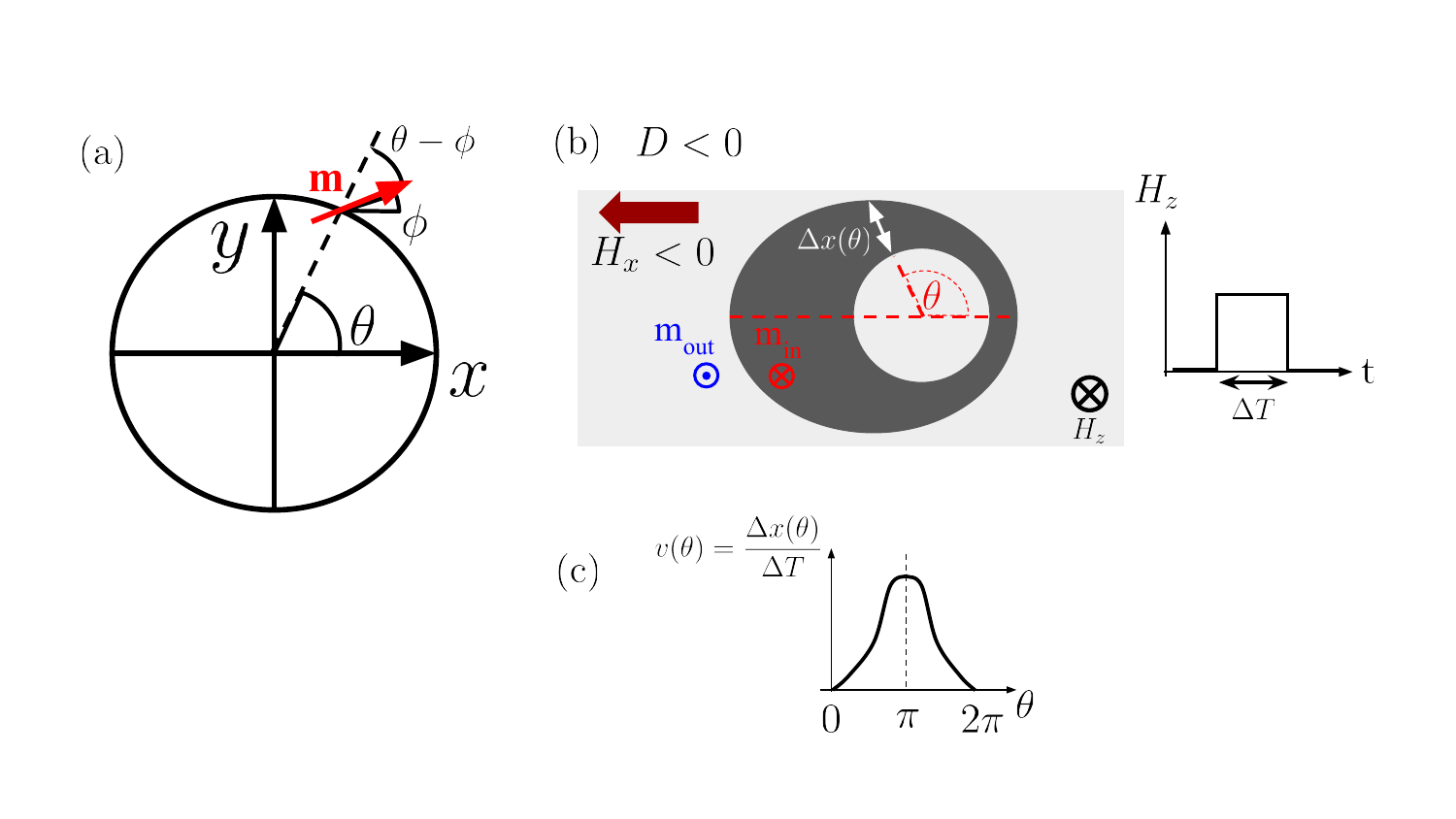}
  \caption{a) Schematic picture of a magnetic bubble domain: $\theta$ represents the angle that parametrizes the
    location on the bubble edge. $\phi$ represents the in-plane magnetization angle in the middle of the DW, i.e. where $m_z = 0$. b) Schematic picture showing the asymmetric bubble expansion with negative DMI ($D<0$) corresponding to counter clockwise rotation of the magnetization. The internal white circle represents the initially nucleated bubble, and the darker gray elliptic shape represents the magnetic bubble expansion driven by an out-of-plane magnetic field pulse in the presence of a DC in-plane field $H_x$ ($H_x < 0$ in the figure). What is shown is the subtraction of the expanded and the initially nucleated bubbles.  The velocity $v(\theta)$ is obtained by dividing the measured displacement $\Delta x(\theta)$ by the (ideally) square pulse duration $\Delta T$ schematically portrayed on the right. The angle $\theta$ is measured from the center of the initially nucleated bubble. c) Schematic velocity profile measured from a single bubble expansion image.}
    \label{fig:bubble_expansion}
\end{figure}

We consider a circular magnetic bubble domain expanding under the combined action of an out-of-plane driving field $H_z$ and an in-plane field $H_x$. The local orientation of the domain wall is described by the angle $\theta$ between the outward wall normal and $H_x$, while the internal magnetization of the wall is described by the angle $\phi$ (see Fig.\ref{fig:bubble_expansion}). The domain-wall energy density $\sigma$ is written as \cite{pakam_anisotropic_2024,thiaville_domain_2002}
\begin{equation}
\begin{split}
\sigma(H_x,\theta,\phi) =
&\,4\sqrt{A K_{\mathrm{eff}}}  \\
&-\pi M_s \Delta
\left[
H_{DMI} \cos(\theta-\phi)
+
H_x \cos\phi
\right] \\
&+\frac{\ln 2}{\pi} t \mu_0 M_s^2
\cos^2(\theta-\phi),
\end{split}
\label{eq:sigma}
\end{equation}
where $\Delta=\sqrt{A/K_{\mathrm{eff}}}$ is the domain-wall width parameter, $A$ is the exchange stiffness, $K_{\mathrm{eff}} = K_u - \frac{\mu_0 M_s^2}{2}$ is the effective perpendicular anisotropy, $M_s$ is the saturation magnetization, $H_{DMI}$ is the DMI field, and $t$ is the effective magnetic thickness entering the magnetostatic wall-anisotropy term. $K_u$ represents the uniaxial magnetocrystalline anisotropy. In the following, we assume that the bubble domain is large enough as to avoid the influence of curvature induced effects \cite{di_pietro_effect_2025, zhang_domain-wall_2018}.
For each wall orientation $\theta$, the equilibrium internal angle $\phi_{eq}$ is obtained by minimizing Eq.~\eqref{eq:sigma},
\begin{equation}
\left.
\frac{\partial \sigma(H_x,\theta,\phi)}{\partial \phi}
\right|_{\phi=\phi_{\mathrm{eq}}}=0 .
\label{eq:phieq}
\end{equation}

In the creep regime, the velocity is exponentially sensitive to the elastic energy controlling thermally activated wall depinning \cite{Lemerle1998, chauve_creep_2000}. Following previous one-dimensional descriptions of Dzyaloshinskii domain-wall creep, the relevant elastic quantity is not only the wall energy, but the relaxed domain-wall stiffness, which includes the energy landscape and the possible relaxation of the internal magnetization \cite{pellegren_dispersive_2017,lau_disentangling_2018}. We therefore define
\begin{equation}
\tilde{\sigma}(\theta,H_x)
=
\sigma
+
\sigma_{\theta\theta}
-
\zeta\bigg(\frac{L}{2 \Lambda}\bigg)
\frac{\sigma_{\theta\phi}^2}{\sigma_{\phi\phi}},
\label{eq:stiffness}
\end{equation}
where all derivatives are evaluated at $\phi=\phi_{\mathrm{eq}}(\theta,H_x)$. The first two terms correspond to the conventional orientational stiffness, while the last term accounts for relaxation of $\phi$ along a finite wall segment of length $L$. $\Lambda = \Delta\sqrt{\frac{\sigma_0}{\sigma_{\phi \phi}}}$ is the length scale for exchange along the domain wall and the function $\zeta(l) = 
 1 - \frac{3}{l^3}(l - \tanh(l))$ interpolates the limits $ \zeta\rightarrow0 ~(l\rightarrow \infty)$ and $ \zeta\rightarrow1 ~ (l \rightarrow 0 )$ \cite{pellegren_dispersive_2017}. In the numerical analysis, we also compare Eq.~\eqref{eq:stiffness} with the limiting case $\tilde{\sigma}=\sigma$, referred to here as the ''stiffness off'' model. This comparison allows the clear separation of the influence of wall stiffness on the angular velocity profile. Chiral damping effects are incorporated through an orientation-dependent effective damping \cite{akosa_phenomenology_2016,jue_chiral_2016}, 
\begin{equation}
\alpha_{\mathrm{eff}}(\theta,H_x)
=
\alpha_G\big(1+\alpha_{\mathrm{cd}}
\cos\!\left[
\phi_{\mathrm{eq}}(\theta,H_x)-\theta
\right]\big),
\label{eq:alphaeff}
\end{equation}
where $\alpha_G$ represents ordinary isotropic Gilbert damping and $\vert\alpha_{\mathrm{cd}}\vert < 1$ quantifies the strength and sign of the chiral damping contribution \cite{akosa_phenomenology_2016, akosa_decoupling_2024, lau_disentangling_2018}. 
To describe angular bubble expansion, we use a reference-normalized form of the creep velocity,
\begin{equation}
\frac{v(\theta,H_x)}{v(\theta_{\mathrm{ref}},H_x)}
=
\frac{\alpha_{\mathrm{eff}}(\theta_{\mathrm{ref}},H_x)}
{\alpha_{\mathrm{eff}}(\theta,H_x)}
\left[
\frac{\tilde{\sigma}(\theta,H_x)}
{\tilde{\sigma}(\theta_{\mathrm{ref}},H_x)}
\right]^{\beta}
\exp
\left\{
-\chi_0 |H_z|^{-1/4}
\left[
r(\theta)^{1/4}
-
r(\theta_{\mathrm{ref}})^{1/4}
\right]
\right\},
\label{eq:vtheta}
\end{equation}
with
\begin{equation}
r(\theta)
=
\frac{\tilde{\sigma}(\theta,H_x)}
{\tilde{\sigma}(\theta,0)} .
\label{eq:rtheta}
\end{equation}
Taking the logarithm of Eq.~\eqref{eq:vtheta} provides an algebraic form that more clearly separates the different contributions:
\begin{equation}
\begin{split}
\ln\!\left[
\frac{v(\theta,H_x)}{v(\theta_{\mathrm{ref}},H_x)}
\right]
=&
\ln\!\left[
\frac{\alpha_{\mathrm{eff}}(\theta_{\mathrm{ref}},H_x)}
{\alpha_{\mathrm{eff}}(\theta,H_x)}
\right] \\
&+\beta
\ln\!\left[
\frac{\tilde{\sigma}(\theta,H_x)}
{\tilde{\sigma}(\theta_{\mathrm{ref}},H_x)}
\right] \\
&-\chi_0 |H_z|^{-1/4}
\left[
r(\theta)^{1/4}
-
r(\theta_{\mathrm{ref}})^{1/4}
\right].
\end{split}
\label{eq:logvtheta}
\end{equation}
The first term isolates the chiral-damping contribution (what we refer to as the ``chiral damping term''), the second is the domain wall energy dependent prefactor associated with the $\beta$ exponent \cite{akosa_phenomenology_2016} (we refer to this term as the ``$\beta$ prefactor'' in the following), finally the third corresponds to the stiffness-dependent creep activation barrier (what we refer to as ``creep exponential'' in the following).
Here, $\chi_0$ is the creep scaling constant, $\beta$  is a coefficient that describes the scaling law between the critical force $H_c$ and the domain wall energy \cite{akosa_phenomenology_2016}, and $\theta_{\mathrm{ref}}$ is an arbitrary reference direction used to remove the unknown absolute velocity scale \cite{akosa_phenomenology_2016}. The formulation in Eq.~\eqref{eq:logvtheta} therefore separates, within a single angular model, the direct contribution of chiral damping from the elastic-stiffness contribution to the creep velocity.

\section{Results and discussion}

We evaluate Eq.~\eqref{eq:vtheta} to determine how stiffness-related and prefactor-related terms modify the angular velocity profile of an expanding magnetic bubble. Figure~\ref{fig:bubble_expansion} shows the geometry, including the angular coordinate $\theta$, the internal wall magnetization $\phi$, and the applied field directions. In the following, velocities are shown as either as $v(\theta)/v(\theta_{\mathrm{ref}})$ or $\ln[v(\theta)/v(\theta_{\mathrm{ref}})]$ as to emphasize the angular profile independently of the absolute creep velocity scale \cite{akosa_decoupling_2024}. Importantly, the purpose of this analysis is not to extract the DMI field from a single domain expansion image, but to assess whether angular creep velocity profiles can be made consistent with DMI values obtained independently from flow-regime and Brillouin light scattering measurements.

\begin{table}[htbp]
 \centering
 \caption{Layer structure of Co-based samples with Pt heavy metal (HM) underlayers}
  \begin{tabular}{@{}lccc@{}}
    \hline
    Sample & FM layer & Bottom layer & Top layer \\
           & (nm)     & (nm)         & (nm)      \\
    \hline
    a$_4$  & Co(0.8)  & Ta(5)/Pt(3)  & Ir(1)/Ta(3) \\
    a$_5$  & Co(0.8)  & Ta(5)/Pt(3)  & Ta(3)       \\
    \hline
  \end{tabular}
  \label{tab:layer_structure_a4_a5}
\end{table}

The logarithmic form of Eq.~\eqref{eq:logvtheta} separates the different contributions to the normalized velocity. Under typical creep conditions, the term proportional to $\chi_0 |H_z|^{-1/4}$ dominates the angular dependence, since $\chi_0$ can be of order $10$--$20$. As a result, the algebraic prefactor governed by $\beta$ generally has a smaller effect on single-image velocity profiles. This hierarchy is not universal: for systems or field regimes with small $\chi_0$, the exponential, algebraic, and chiral-damping contributions become more comparable, and prefactor effects can become much more pronounced. Some notable cases of small values of $\chi_0$ can be found, for example, in thin films exposed to helium ion irradiation or thermal annealing \cite{Magni2021}.

\begin{table}
 \centering
 \caption{Magnetic properties of Pt/Co based layers. Details on the experimental procedures to obtain the values are reported in \cite{magni_key_2022}. The effective DMI field $H_{DMI}$ is obtained using asymmetric bubble expansion in the flow regime \cite{ha_pham_very_2016}, $D_{\text{Flow}}$ is obtained via Walker field measurements \cite{ha_pham_very_2016}. Both approaches are described in the Supporting information section \ref{sec:Supporting}. $D_{BLS}$ is independently obtained via Brillouin light scattering (BLS) measurements described in \cite{magni_key_2022}.}
  \resizebox{\linewidth}{!}{%
  \begin{tabular}{@{}lcccccccc@{}}
    \hline
    Sample &  $M_s$
  & $K_{\mathrm{eff}}$
  & $A^{\mathrm{a}}$
  & $\chi_0$ 
  &$\alpha_{CD}$
  &$\mu_0 H_{DMI}$ 
  &$D_{\mathrm{Flow}}$ 
  &$D_{\mathrm{BLS}}$ \\
  & (MA/m) & (MJ/m$^3$)
  & (pJ/m) & (T$^{1/4}$) & (-)& (T)&(mJ/m$^2$)&~(mJ/m$^2$)\\
    \hline
    a$_4$  & 1.20 $\pm$ 0.09  & 0.513 $\pm$ 0.002& 12 $\pm$ 3& 12.87 $\pm$ 0.01 & 0.8&  -0.088 $\pm 0.008$ & -0.52 $\pm 0.05$ &-0.56 $\pm$ 0.01 \\
    a$_5$  & 0.90 $\pm$ 0.02   & 0.563 $\pm$ 0.005& 22 $\pm$ 6& 13.42 $\pm$ 0.02  & 0.5&  -0.16 $\pm 0.016$ &-0.91 $\pm 0.09$ &-0.81 $\pm$ 0.01 \\
    \hline
  \end{tabular}%
  }

  \vspace{0.5em}
  \begin{flushleft}
  \footnotesize $^{\mathrm{a}}$ Exchange stiffness values were obtained through the complementary measurement technique described in the Appendix.
  \end{flushleft}

  \label{tab:experimental}
\end{table}

\begin{figure}
    \centering
    \includegraphics[width=0.75\linewidth]{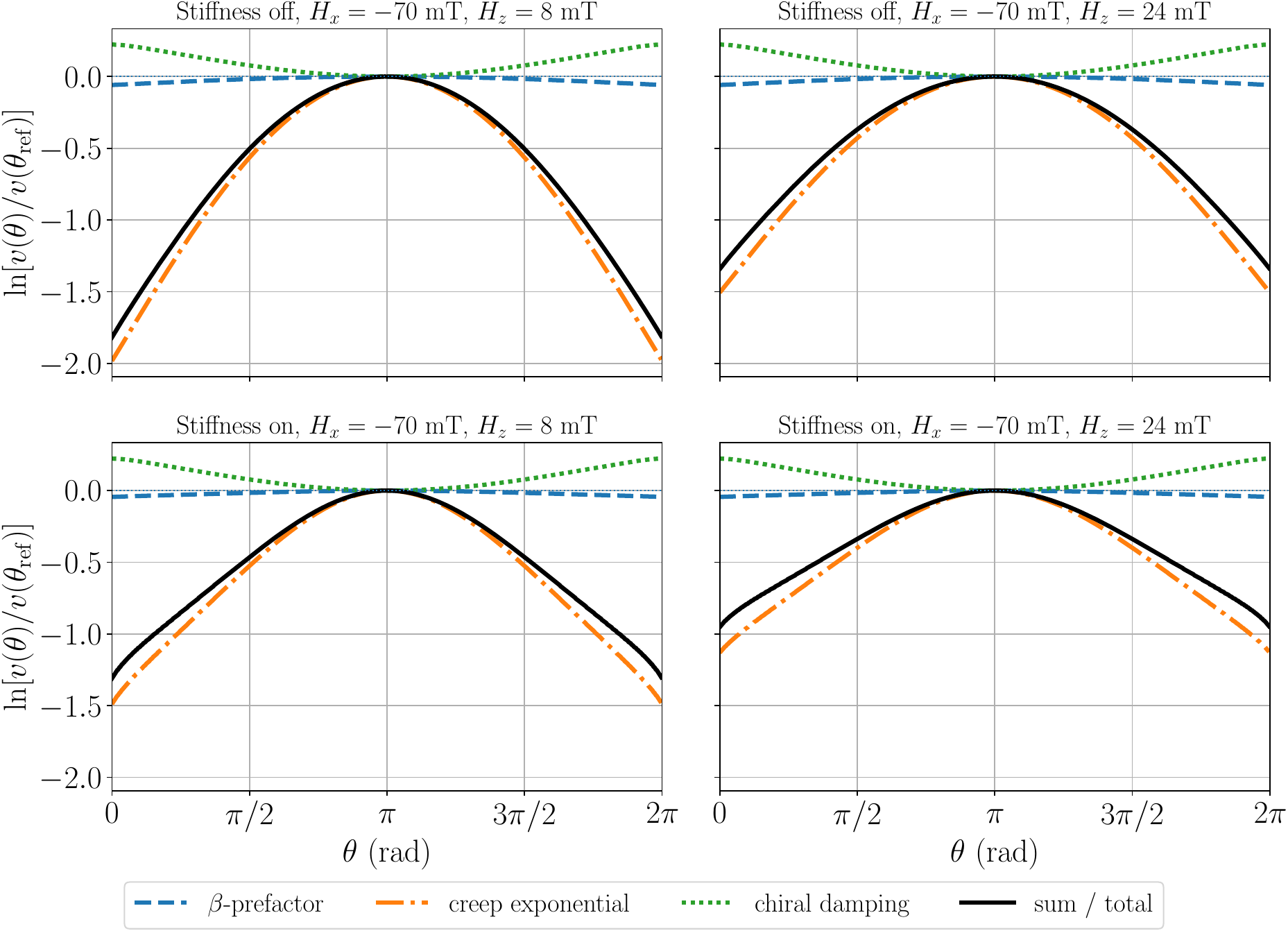}
    \caption{
Logarithmic decomposition of the normalized angular velocity,
$\ln[v(\theta)/v(\theta_{\mathrm{ref}})]$, for $H_x=-70~\mathrm{mT}$ and $\theta_{\mathrm{ref}}=\pi$.
The four panels compare the contributions with domain-wall stiffness switched off (top row) and on (bottom row), for two out-of-plane driving fields, $H_z=8~\mathrm{mT}$ and $H_z=24~\mathrm{mT}$.
The individual terms correspond to the algebraic prefactor, the creep exponential, and the chiral-damping contribution, while the black curve shows their sum. The magnetic parameters used refer to sample $a_5$, i.e. the second column of Tab.~\ref{tab:experimental}. For the ``stiffness on'' variants, a value of $L = 5$ nm is chosen.
}
    \label{fig:CD_stiffness}
\end{figure}

As a relevant case study of Eq.~\eqref{eq:logvtheta}, we consider magnetic parameters describing Pt/Co-based ultrathin samples, 
 called hereafter $a_4 $ and $a_5$, prepared by magnetron sputtering at the University of Leeds (see Tables \ref{tab:layer_structure_a4_a5} and \ref{tab:experimental} as well as \cite{magni_key_2022} for details).
 Figure~\ref{fig:CD_stiffness} illustrates how the different terms in the model contribute to the normalized angular velocity. The magnetic parameters used to calculate the curves of Fig \ref{fig:CD_stiffness}, are related to sample $a_5$. From the bottom panels of Fig \ref{fig:CD_stiffness}, we observe how including the relaxed stiffness $\tilde{\sigma}(\theta)$ produces a strong quantitative modification of $v(\theta)$ because it changes the activation barrier entering the creep exponent. Chiral damping also modifies the curve, but it mainly enters as a prefactor-related angular correction. Therefore, the velocity profile $v(\theta)$ does not generally exhibit a feature that can be assigned uniquely to chiral damping alone. Even when stiffness is included, chiral damping appears primarily as a quantitative reshaping of the angular curve rather than as an unambiguous qualitative signature. The log-decomposition plots (Fig.\ref{fig:CD_stiffness}) therefore show that, although the model terms are mathematically separable, their experimental identification from $v(\theta)$ alone remains challenging; subtle differences may be obscured by contour extraction, normalization, and material-parameter uncertainties.

\begin{figure}
    \centering
    \includegraphics[width=\linewidth]{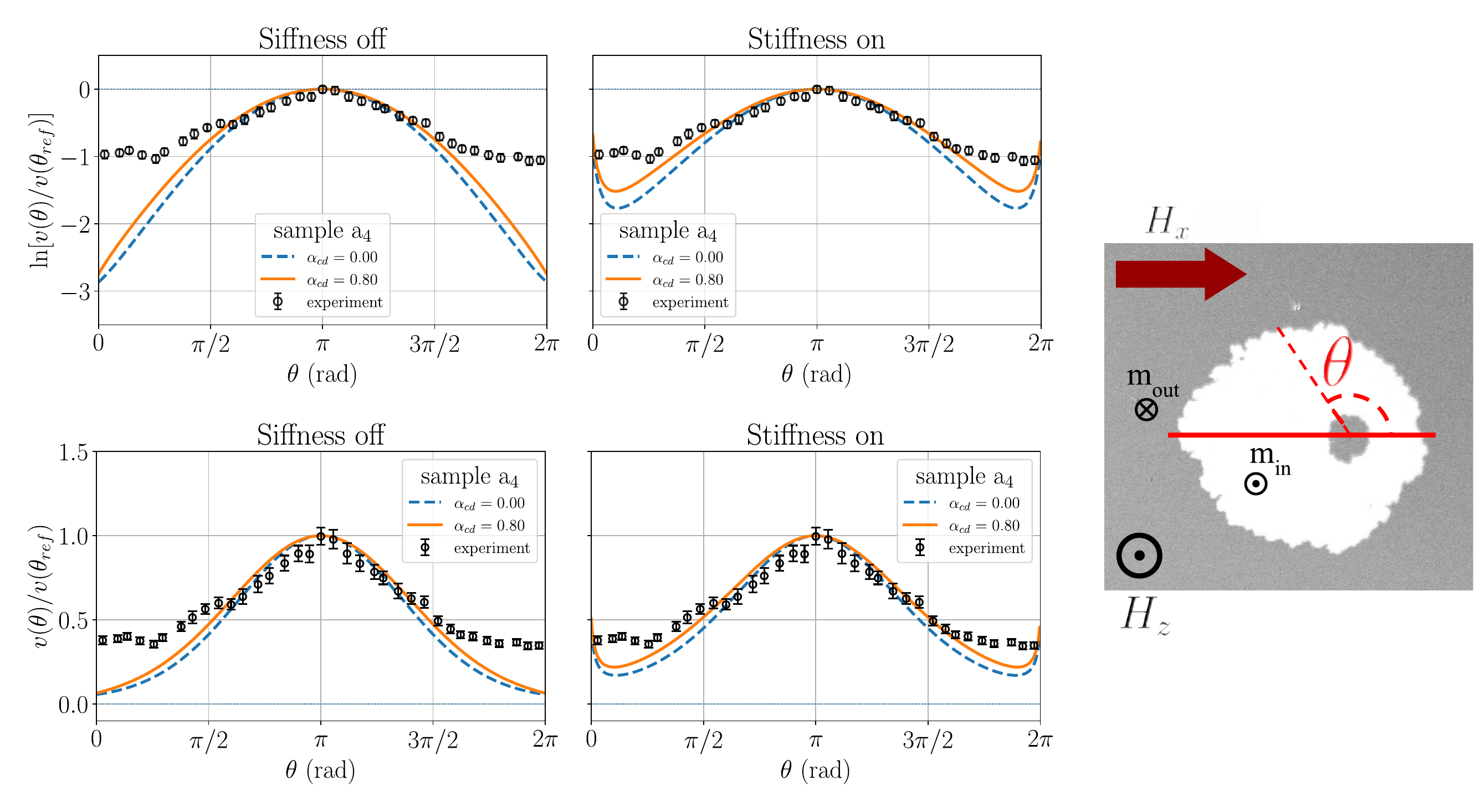}
    \caption{
Angular velocity profile for sample $a_4$ at $H_z=8.8~\mathrm{mT}$, in the presence of $H_x=60~\mathrm{mT}$, plotted as $\ln[v(\theta)/v(\theta_{\mathrm{ref}})]$ on the top row and as $v(\theta)/v(\theta_{\mathrm{ref}})$ on the bottom row. In both cases we set  $\theta_{\mathrm{ref}}=\pi$. Open symbols show the experimental profile, while lines show model calculations for four cases: energy-only, energy-only with chiral damping, stiffness-only, and the full model including both relaxed domain-wall stiffness and chiral damping. The DMI field is fixed to the independently determined value $\mu_0 H_{DMI} = -88$ mT reported in Tab.~\ref{tab:experimental}. In the ``stiffness on" case, we set $L=90$ nm (see Eq.\eqref{eq:stiffness}). On the right, we show the experimental differential Kerr image from which $v(\theta)$ is computed. The clear contrast represents magnetization OOP. 
}
    \label{fig:a4_vtheta_comparison}
\end{figure}

\begin{figure}
    \centering
    \includegraphics[width=\linewidth]{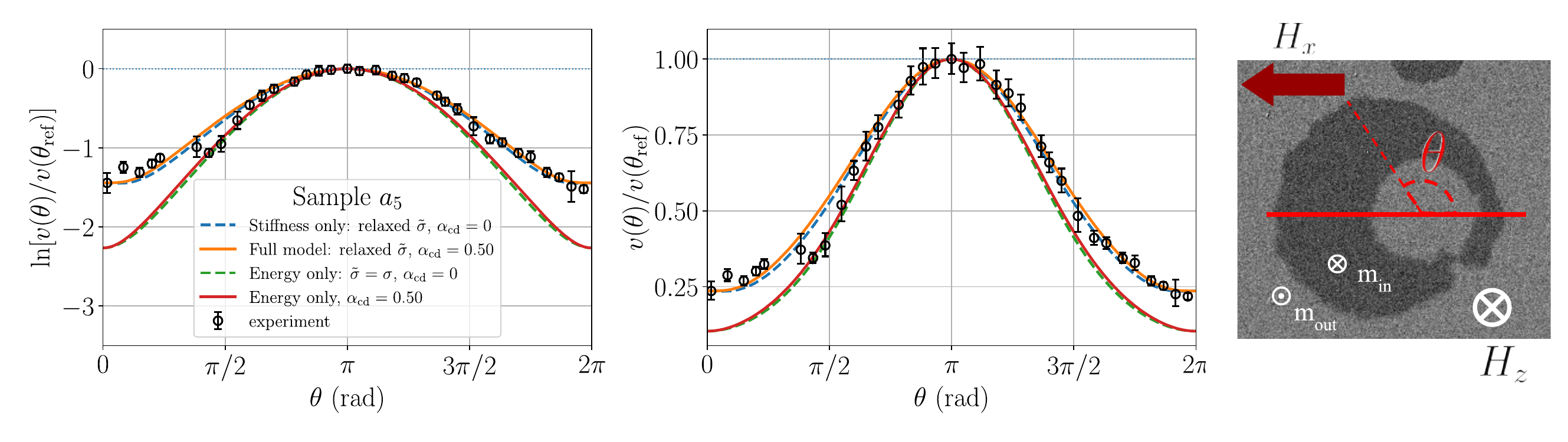}
    \caption{
Angular velocity profile for sample $a_5$ at $H_x=-80~\mathrm{mT}$ and $H_z=-21.21~\mathrm{mT}$, plotted as $\ln[v(\theta)/v(\theta_{\mathrm{ref}})]$ (left panel) and $v(\theta)/v(\theta_{\mathrm{ref}})$ on the right panel. In both cases we set  $\theta_{\mathrm{ref}}=\pi$. Open symbols show the experimental profile, while lines show model calculations for four cases: energy-only, energy-only with chiral damping, stiffness-only, and the full model including both relaxed domain-wall stiffness and chiral damping. The DMI field is fixed to the independently determined value $\mu_0 H_{DMI} = -160$ mT reported in Tab.~\ref{tab:experimental}. In the ``stiffness on" case, we set $L=50$ nm (see Eq.\eqref{eq:stiffness}). The data is taken from the Kerr image shown on the right. 
}
    \label{fig:a5_vtheta_comparison}
\end{figure}

The relevance of these terms becomes most evident when considering the consistency of DMI values across different measurement techniques. Figures~\ref{fig:a4_vtheta_comparison} and~\ref{fig:a5_vtheta_comparison} show representative experimental DW velocity angular profiles for samples $a_4$ and $a_5$; the DMI fields for these samples are determined by asymmetric DW propagation in the flow regime \cite{ha_pham_very_2016} and cross-checked against Brillouin light scattering measurements (see \cite{magni_key_2022} for BLS measurements and Sec.\ref{sec:Supporting} for flow measurements). The error bars shown in Figs \ref{fig:a4_vtheta_comparison} and \ref{fig:a5_vtheta_comparison} are obtained by coarse graining the  measurements into finite angular bins of width $\Delta \theta$; for each bin, the plotted point corresponds to the mean angle and mean normalized velocity, while the error bar represents the standard error of the mean of the velocity values within that bin. 

Using these DMI values, an energy-only angular creep model does not reproduce the measured angular velocity profiles (see Figs \ref{fig:a4_vtheta_comparison}-\ref{fig:a5_vtheta_comparison}
). This description has the tendency to vastly overestimate the effect of $H_{DMI}$, predicting a much stronger discrepancy between the fast part of the bubble (i.e. $v(\theta = \pi)$) and the slow part of the bubble (i.e. $v(\theta = 0)$). When relaxed domain-wall stiffness and chiral damping are included, the measured profiles can be reproduced while keeping DMI values consistent with the independent flow and BLS measurements. Sharp kinks may appear in the calculated angular velocity when the relaxed domain-wall stiffness is included (see e.g. Fig.\ref{fig:a4_vtheta_comparison}). These features arise because $\tilde{\sigma}$ contains angular-curvature terms and near the crossover between mixed Bloch-N\'eel and saturated N\'eel configurations, small changes in $\phi_{\rm eq}$ can produce cusp-like variations in $\tilde{\sigma}$, which are further amplified by the creep exponential. In experiments, however, such features are expected to be strongly rounded by disorder, thermal averaging, finite imaging resolution, and contour-extraction procedures; their absence as sharp kinks in measured profiles should therefore not be taken as evidence against the stiffness contribution.

This result  illustrates that  fitting experimental curves with the energy-only angular creep model \cite{Je2013} may lead to DMI values different - in this case smaller - from those measured experimentally with more reliable, independent methods. The trend of underestimating DMI via DW motion measurements in the creep regime has been reported in several samples \cite{magni_datasets_2022,soucaille_probing_2016} and will be treated in depth in future work. Here, we show that the full model of Eq.~\eqref{eq:logvtheta} can address this apparent discrepancy by showing that the angular creep profile is sensitive to additional terms beyond the wall energy alone. However, the description of Eq.\eqref{eq:logvtheta} introduces additional parameter ambiguity, namely chiral damping strength $\alpha_{CD}$, and the stiffness relaxation length  $L$, which are not independently constrained by the angular profile. Therefore, in the absence of independent measures or estimates for these two physical quantities, $v(\theta)$ should not be viewed as a stand-alone route to determine $H_{DMI}$. Importantly, the used values $L=90$ nm and $L=50$ nm  for samples $a_4$ and $a_5$, respectively, should be regarded as effective relaxation lengths. Their difference is qualitatively consistent with a longer internal-angle relaxation scale in the sample with lower $A$ and slightly lower $K_{\mathrm{eff}}$, but $L$ may also absorb extrinsic effects such as disorder, pinning length, bubble curvature, and contour-extraction averaging \cite{pellegren_dispersive_2017}. We therefore use $L$ as a phenomenological parameter controlling the degree of stiffness relaxation, rather than as a directly measured material constant.

Our conclusion is that a rigorous determination of DMI field from creep-regime velocity profiles requires prior or complementary information on chiral damping and domain-wall stiffness. Recent studies of BLS spectra suggest that chiral damping can be measured independently from the asymmetric broadening of Stokes and anti-Stokes peaks \cite{kim_chiral_2024}. By contrast, an independent experimental measurement of the domain-wall stiffness relevant to creep dynamics is, to our knowledge, still missing and will be addressed in future work.

Overall, domain-wall stiffness and chiral damping can strongly affect angular bubble expansion, even when their individual signatures are not visually separable in the velocity profile. Stiffness is particularly important under conventional creep conditions because it modifies the dominant exponential term, whereas prefactor-related effects become more relevant for small $\chi_0$ or strong chiral damping. The extended model therefore highlights the importance of elastic effects and chiral damping for the interpretation of DMI metrology and for the design of velocity-selective DW logic or DW memory devices.

\section{Conclusion}

In conclusion, we have extended the angular creep description of magnetic bubble expansion by incorporating domain-wall stiffness and chiral damping into the normalized velocity profile $v(\theta)$ in the presence of an applied in-plane field. The method is compatible with fast, single-image measurements of asymmetric domain-wall velocity, but the analysis shows that such images do not by themselves provide an unambiguous DMI field value. Instead, when DMI fields obtained independently from flow-regime and BLS measurements are imposed, reproducing the corresponding creep-regime angular profiles requires going beyond energy-only models. Domain-wall stiffness can strongly modify the angular dependence through the creep activation barrier, while chiral damping enters through the prefactor and may not produce a unique signature in the velocity profile alone. These results highlight the importance of combining angular velocity analysis with independent constraints on DMI, chiral damping, and domain-wall stiffness, providing a more reliable framework for interpreting chiral domain-wall dynamics.

\appendix
\section{Flow-regime measurements}
\label{sec:Supporting}

\begin{figure}
    \centering
    \includegraphics[width=0.75\linewidth]{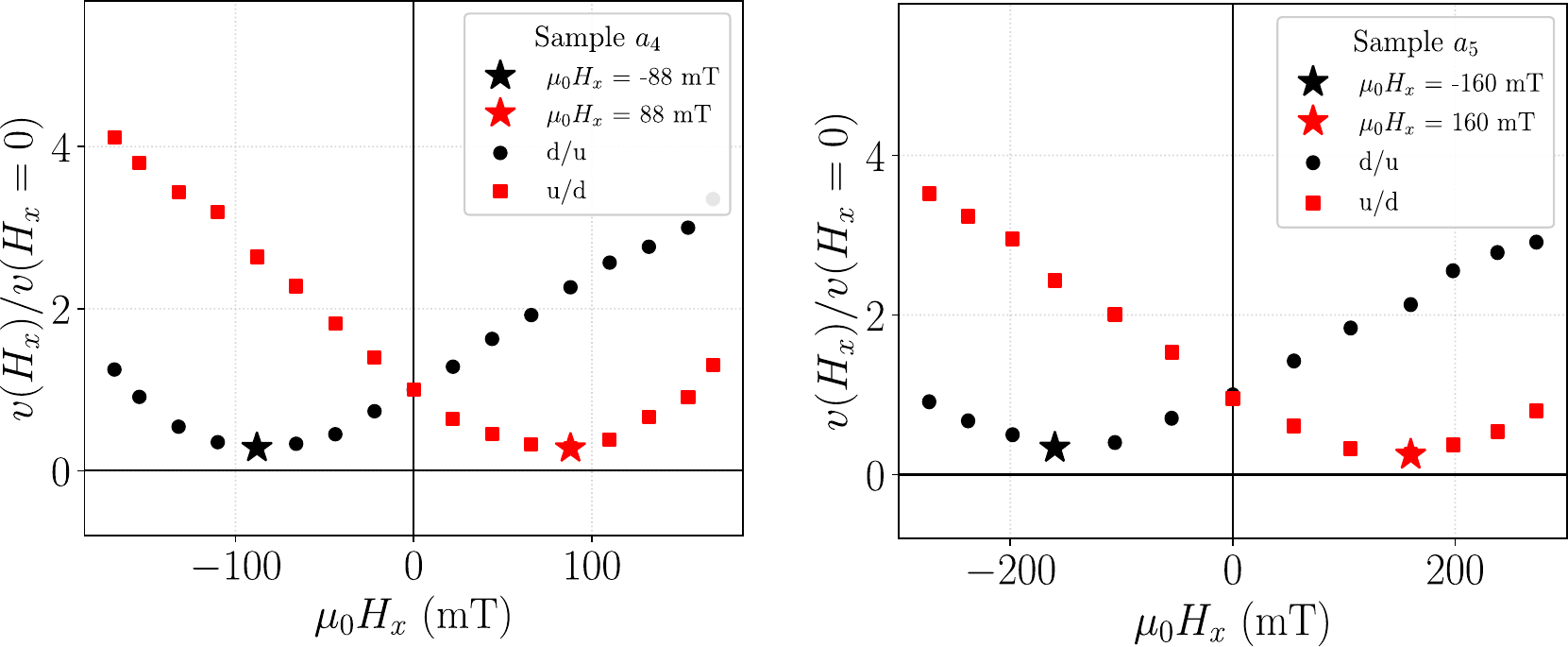}
    \caption{Asymmetric bubble expansion measurements as a function of $\mu_0 H_x$ performed in the flow regime for the up/down (``u/d") and down/up (``d/u") portions of the bubble domain. The location of the velocity minimum corresponds to $\mu_0 H_x = -\mu_0 H_{DMI}$ \cite{krizakova_study_2019}.}
    \label{fig:Flowparabs_a4_a5}
\end{figure}

In the flow regime, asymmetric bubble expansion is not affected by the creep activation barrier \cite{metaxas_creep_2007} and the dynamics of the chiral DWs can effectively be described by 2D model (\cite{ha_pham_very_2016}). The position of the velocity minimum as a function of in-plane field can be used to estimate the compensation field at which the applied field balances the DMI effective field\cite{ha_pham_very_2016,krizakova_study_2019,garcia_magnetic_2021}. Figure~\ref{fig:Flowparabs_a4_a5} shows the corresponding measurements for samples $a_4$ and $a_5$, where the minima are clearly visible. The extracted $H_{DMI}$ values, are reported in Table~\ref{tab:Aex}. 

\begin{figure}
    \centering
    \includegraphics[width=0.5\linewidth]{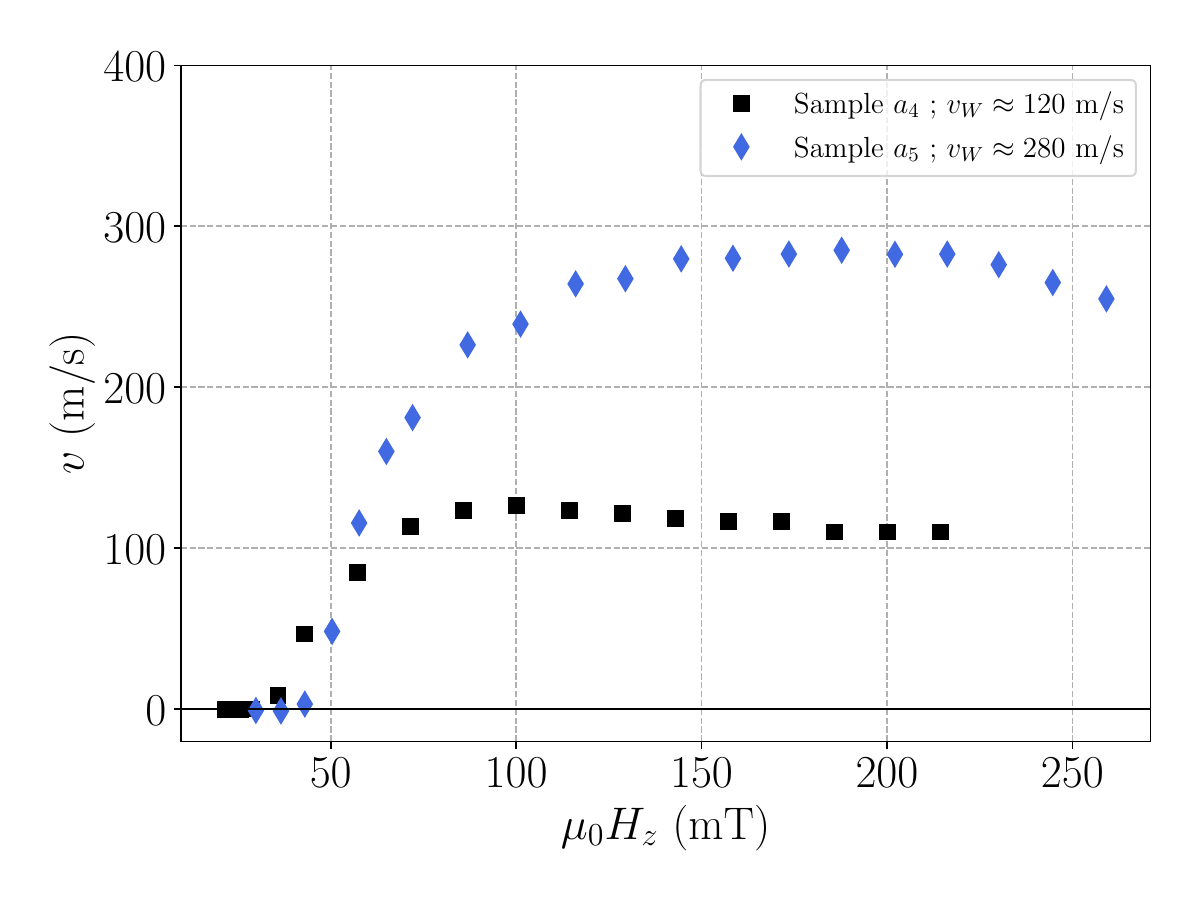}
    \caption{Domain-wall velocity measured as a function of out-of-plane field $\mu_0 H_z$ for samples $a_4$ and $a_5$ (see Table~\ref{tab:layer_structure_a4_a5}). The plateau velocity visible in the curve is then used to compute $D$ via Eq.\eqref{eq:plateau}.}
    \label{fig:a4_a5_WB}
\end{figure}

A second, independent estimate of the DMI strength can be obtained from the plateau velocity observed after the Walker field in samples with sufficiently large DMI constants, providing access to its value via the expression \cite{ha_pham_very_2016,garcia_magnetic_2021}

\begin{equation}
    v_W = \frac{\pi}{2}\gamma\frac{D}{M_s}, 
    \label{eq:plateau}
\end{equation}

where $v_W$ is the Walker velocity, $\gamma$ is the gyromagnetic ratio, and $M_s$ is the saturation magnetization. Figure~\ref{fig:a4_a5_WB} shows the measurements used to identify the Walker velocity for samples $a_4$ and $a_5$. The resulting DMI constants are reported in Table~\ref{tab:Aex}. These values are compatible with the independently measured BLS values reported in Table~\ref{tab:experimental}, supporting the DMI fields used in the main analysis.

The two flow-regime measurements therefore provide complementary DMI quantities: the DMI field $H_{DMI}$ from the in-plane-field compensation condition and the DMI constant $D$ from the Walker velocity. As a useful consistency check, these two quantities can also be combined to estimate the exchange stiffness. Since

\begin{equation}
D = \mu_0 M_s H_{DMI} \Delta ,
\label{eq:D_HD_relation}
\end{equation}

with

\begin{equation}
\Delta = \sqrt{\frac{A}{K_\mathrm{eff}}},
\end{equation}

one obtains

\begin{equation}
A =
K_\mathrm{eff}
\left(
\frac{D}{\mu_0 M_s H_{DMI}}
\right)^2 .
\label{eq:A_from_flow}
\end{equation}

This estimate of $A$ is not the primary purpose of the flow measurements, but it provides an independent consistency check on the material parameters entering the angular creep calculation. It is particularly useful because direct measurements of the exchange stiffness in ultrathin multilayers are experimentally challenging. Conventional approaches include extracting $A$ from the temperature dependence of the saturation magnetization, using Bloch-law or related spin-wave analyses of $M_s(T)$ \cite{yastremsky_thermodynamics_2019,kuzmin_exchange_2020}, or from spin-wave spectra measured by ferromagnetic resonance, perpendicular standing spin-wave modes, or Brillouin light scattering \cite{devolder_exchange_2016,waring_exchange_2023}. In ultrathin multilayers, these approaches can be complicated by interface anisotropy, magnetic dead layers, finite-size effects, and uncertain exchange boundary conditions.

Applying Eq.~\eqref{eq:A_from_flow} to samples $a_4$ and $a_5$ gives the exchange stiffness values reported in Table~\ref{tab:Aex}.

\begin{table}[htbp]
 \centering
 \caption{$H_{DMI}$, $D$, and $A$ obtained from domain-wall velocity measurements in the flow regime as described in \cite{garcia_magnetic_2021}. Since both $\mu_0 H_{DMI}$ and $D$ are extracted visually from plots, rather than from a numerical fitting procedure, a 10\% uncertainty is assumed on both $\mu_0 H_{DMI}$ and $D$. The uncertainty on $A$ also includes the propagated uncertainties on $M_s$ and $K_{\mathrm{eff}}$ from Table~\ref{tab:experimental}.}
  \begin{tabular}{@{}lccc@{}}
    \hline
    Sample & $\mu_0 H_{DMI}$ (Flow)& $D$ (Flow)& $A$\\
           & T& mJ/m$^2$& pJ/m\\
    \hline
    a$_4$  & $-0.088 \pm 0.008$& $-0.52 \pm 0.05$& $12 \pm 3$\\
    a$_5$  & $-0.160 \pm 0.016$& $-0.91 \pm 0.09$& $22 \pm 5$\\
    \hline
  \end{tabular}
  \label{tab:Aex}
\end{table}
\begin{acknowledgments}
This project is supported by the European Union (EU) and the Italian ministry of University and Research (MUR) under the grant PRIN 2022 ``Metrology for spintronics: A machine learning approach for the reliable determination of the Dzyaloshinskii--Moriya interaction (MetroSpin)'' Codice di progetto: 2022SAYARY. Yasser Shokr gratefully acknowledges financial support from T{\"U}B{\.I}TAK under
grant number 124C509, which supported his work. The work of Stefania Pizzini was supported by the France 2030 government plan
managed by the Agence Nationale de la Recherche (project PEPR SPIN CHIREX ANR-22-EXSP-0002). This work was supported in part by the European Metrology Programme for Innovation and Research (EMPIR) Programme co-financed by the Participating States under Project 17FUN08-TOPS.
\end{acknowledgments}

\medskip

%


\end{document}